\documentstyle[preprint,prc,tighten,eqsecnum,aps]{revtex}
\begin{document}
\draft
\preprint {McGill/94--17.}
\title{\bf Momentum--dependent nuclear mean fields and collective flow\\ in
heavy
ion collisions}
\author{Jianming Zhang\thanks{email: jzhang@hep.physics.mcgill.ca},
Subal Das Gupta\thanks{email: dasgupta@hep.physics.mcgill.ca}, and Charles
Gale\thanks{email: gale@hep.physics.mcgill.ca}}
\vspace{1cm}
\address{Physics Department, McGill University \\ 3600 University St.,
Montr\'eal
QC, Canada H3A--2T8}
%\begin{instit}
%Physics Department
%\end{instit}
%\begin{instit}
%McGill University, Montr\'eal, Qu\'ebec, Canada H3A--2T8
%\end{instit}
%\receipt{---------1994  }

\maketitle

\begin{abstract}
 We use the Boltzmann-Uehling-Uhlenbeck model to simulate the dynamical
evolution of heavy ion collisions and to compare the effects of two
parametrizations of the
momentum--dependent nuclear mean field that have identical properties in cold
nuclear matter. We compare with recent data on nuclear flow, as characterized
by
transverse momentum distributions and flow ($F$) variables for symmetric and
asymmetric systems.
We find that the precise functional dependence of the nuclear mean field on the
particle momentum is important.
With our approach, we also confirm that the difference
between symmetric and
asymmetric systems can be used to pin down the density and momentum dependence
of the nuclear self consistent one--body potential, independently.
All the data
can be reproduced very well with a momentum--dependent interaction with
compressibility K = 210 MeV.
\end{abstract}
\newpage
%----------- Text -----------
\section{Introduction}
Over the past decade, the extraction of the nuclear equation of state (EOS)
from experimental data has been one of the main goals
of intermediate energy heavy ion collisions. The nuclear EOS plays a
major role in the physics of colliding nuclei at high energies and also
has a
major influence in the theory of supernov{\ae} explosions and neutron star
properties
\cite{grsto}. Information on the EOS, as characterized by its coefficient of
nuclear
compressibility, K, can also be deduced from detailed Hartree--Fock plus RPA
analysis of giant monopole resonances in finite nuclei
\cite{blaizot}.

In the framework of heavy ion collision physics in the 100 MeV/A $\sim$ 2 GeV/A
energy regime  and its relation to the nuclear
EOS, the measurement and theoretical interpretation  of collective flow
observables have been vital \cite{hhg89}. Among the many models suggested to
describe theoretically heavy ion collisions at such energies,
the Boltzmann-Uehling-Uhlenbeck (BUU) approach
model has proven to be very successful \cite{gfb88}. In
BUU simulations, nucleons can suffer hard collisions and can also move on
curved trajectories, owing to interaction with the self--consistent nuclear
mean field. The
properties of the mean field are crucial to such calculations and can also be
directly related to the nuclear equation of state. Some effort has been devoted
to obtain realistic nuclear mean fields that could be used in practice within
such
numerical approaches.

Early on in microscopic analyses, it
appeared that the data on nuclear flow, as characterized by transverse momentum
plots~\cite{pd85,doss86} and flow angle distributions~\cite{hag84} demanded an
equation of state with a high compressibility coefficient
(K $\approx$ 380 MeV) \cite{hs86}. However, it was later shown that if a
reasonable momentum dependence was introduced in the nuclear mean field, a
lower compressibility would be favored in the interpretation of the
experimental data \cite{gale87,ja87,gale90}. In fact, Pan and
Danielewicz~\cite{pan93} have recently shown that flow data for
asymmetric systems could differentiate between a hard momentum--independent EOS
and a
soft momentum--dependent EOS, favoring the latter.
Finally, it is clear that the momentum dependence of the nuclear
mean field is an unavoidable feature for  a fundamental understanding
of nuclear matter properties \cite{jeukenne} and for the successful
interpretation of current heavy ion data.

Additional properties of momentum--dependent mean fields have also emerged in
the BUU analysis of heavy collisions. Different sets of
momentum--dependent parametrizations sharing a common compressibility
coefficient have been used. We will concentrate on two
of those. We label them GBD~\cite{gale87} and MDYI~\cite{gmw88}, in accordance
with the (quoted) articles in which they have been introduced. This
nomenclature has been used previously \cite{gale90}.
Pan and
Danielewicz \cite{pan93} have used a GBD-type parametrization. Another
momentum--dependent
potential used in one-body numerical simulations is associated with the Gogny
interaction \cite{mota92}. The properties of the GBD and MDYI potentials
are somewhat similar in the ground state, but they will have
different behaviors in actual dynamical situations \cite{gmw88,koch92}. We
shall discuss this aspect in the present paper.

Stimulated by the findings of Ref. \cite{pan93}, we have analyzed the
quantitative differences between GBD and MDYI--type approaches. We also give
our
own opinion as to which parametrization should be used in calculations where
nonequilibrium effects can be important, as in intermediate energy heavy ion
collisions. We further explore the impact of our conclusions on the
determination of the nuclear EOS, by comparing with current heavy ion data.
We perform BUU calculations for symmetric and asymmetric projectile--target
combinations, at various colliding energies. We also comment on the
quantitative
importance of angular momentum conservation at the microscopic level in the
interpretation of nuclear transverse momentum data.
Our paper is organized in the following way:
in the next section we give a detailed presentation of the nuclear mean fields
used
in our BUU calculations. The following section analyzes the generation of
transverse momentum. We then compare with experimental data and we finally
conclude.

\section{Nuclear momentum--dependent mean fields}
As mentioned above, different forms of phenomenological
momentum--dependent potentials
are found in BUU applications. Gale, Bertsch and Das Gupta
employed  a parametrization of the potential energy density that can be written
as~\cite{gale87}
\begin{eqnarray}
V_{\rm GBD}\, (\rho ( \vec{r}) )\,=\,\frac{A}{2}\frac{\rho^{2} ( \vec{r} )}
{\rho_{0}}+\frac{B}{\sigma+1}
\frac{\rho^{\sigma+1} ( \vec{r} )}{\rho^{\sigma}_{0}}
+\frac{C \rho (\vec{r} ) }{\rho_{0}}\int\!\,d^{3}p\ {f( \vec{r} ,\vec{p})
\over 1+\left[
{{\vec{p} - < \vec{p} >}\over { \Lambda}}\right]^{2}}
\label{eq:v2}
\end{eqnarray}
%which can be obtained from a skyrme force in lowest
%order~\cite{koch88}.
The corresponding mean field (GBD) is obtained by taking a functional
derivative with respect to the single--particle occupation function; $U\ =
\frac
{\delta V}{\delta f} |_{\vec{p}}$. One then obtains
\begin{equation}
U_{\rm GBD}\, (\rho (\vec{r} ) , \vec{p} )=A \left(\frac{\rho ( \vec{r} )}
{\rho_{0}}\right) +B \left( \frac{\rho (\vec{r} )}
{\rho_{0}}\right)^{\sigma}
+\frac{C}{\rho_{0}}\int\!\,d^{3}p^{'} {f( \vec{r} , \vec{p}^{\,\prime}) \over
1+\left[\frac{ \vec{p}^{\,\prime} - < \vec{p} >}{\Lambda}\right]^{2}}
+\frac{C}{\rho_{0}} \frac{\rho}{1+\left[\frac{\vec{p} - < \vec{p}
>}{\Lambda}\right]^{2}}\ ,
\label{eq:u2}
\end{equation}
where $\vec{p}$ is the momentum of the particle, $<\vec{p}>$ is a local
momentum average, and $f(\vec{r},\vec{p})$ is the phase space occupation
density. This
quantity is normalized such that the nuclear density
$\rho( \vec{r} )$~=~$\int\,d^{3}p~f(
\vec{r},\vec{p} )$.
There are five parameters to be determined in $U_{\rm GBD} (\rho (\vec{r}),
\vec{p})$. Previously
\cite{gale87}, two of them were chosen arbitrarily:
the momentum scale $\Lambda$ = 400 MeV and $\sigma=7/6$. This exponent
has an especially large influence on the nuclear equation of state
compressibility coefficient, K. We further required the following at the
saturation density: (i) the effective mass $m^*/m$ was set to 0.7 at the Fermi
surface, and (ii) the total energy per nucleon was adjusted to reproduce the
volume term of the semi--empirical mass formula, E/A = $-$16 MeV. We used
$\rho_{0}=0.163\ fm^{-3}$ and thus obtained K = 215 MeV.

Some subsequent work by Welke et al. ~\cite{gmw88} used an improved
parametrization
\begin{equation}
V_{\rm MDYI} (\rho ( \vec{r} )=\frac{A}{2}\frac{\rho^{2} ( \vec{r}
)}{\rho_{0}}+\frac{B}{\sigma+1}
\frac{\rho^{\sigma+1} ( \vec{r} )}{\rho^{\sigma}_{0}}
+\frac{C}{\rho_{0}}\int\int\!\,d^{3}p\,d^{3}p' {f( \vec{r}, \vec{p})f(
\vec{r}, \vec{p}^{\,\prime} )
\over 1+\left[\frac{ \vec{p} - \vec{p}'}{\Lambda}\right]^{2}}
\label{eq:v1}
\end{equation}
which leads to the other form of the momentum--dependent potential we shall
consider:
\begin{equation}
U_{\rm MDYI} (\rho (\vec{r}), \vec{p})=A \left(\frac{\rho (\vec{r} )}{\rho_{0}}
\right)+
B \left(\frac{\rho (\vec{r} )}{\rho_{0}}\right)^{\sigma}
+2\ \frac{C}{\rho_{0}}\int\!\,d^{3}p^{'} {f( \vec{r}, \vec{p}^{\,\prime}) \over
1 + \left[\frac{ \vec{p} - \vec{p}^{\,\prime}}{\Lambda}\right]^{2}}\ .
\label{eq:u1}
\end{equation}
The five constants A, B, C, $\sigma$, and $\Lambda$
in $U_{\rm MDYI} (\rho ( \vec{r}), \vec{p} )$ were set by demanding that, at
saturation:
E/A = $-$16 MeV, K = 215 MeV, the real part of the optical potential
$U(\rho_{0},~p~=~0)~=~-75~{\rm MeV}$
and $U(\rho_{0},~\frac{p^{2}}{2m}~=~300~{\rm MeV})~=~0$. It then follows that
$U(\rho_{0},~p\rightarrow~\infty)~=~30.5~{\rm MeV}$ and that the effective
mass $m^*/m$~=~0.67, at the Fermi surface. The agreement of $U_{\rm MDYI}$
with the real
part of the optical potential as
extracted from experiment is remarkable, at both low and high energies
\cite{gale92}.  To clarify the origins of these parametrizations, we state here
that a Yukawa interaction would have a mean field whose exchange
term would be a momentum--dependent expression of the MDYI type \cite{gmw88}.
The GBD potential energy density  can be obtained from its MDYI countrepart
by replacing $\vec{p}^{\,\prime}$ in the denominator of the integrand of Eq.
(\ref{eq:v1}) by its average, $< \vec{p}^{\,\prime} >$. The momentum--dependent
term of the MDYI mean field is
attractive and important at low momentum, but it
weakens and disappears at very high momentum.
Even though both of the above parametrizations (GBD and MDYI) can share the
same
compressibility K, the quantities  $U(\rho_{0},~p\rightarrow~\infty)$ and the
effective mass m* can be different.  The value of
$U(\rho_{0},~p\rightarrow~\infty)$ has
important consequences for the modelling of nuclear collisions at high
energies, as we shall see.

In this work, for the sake of consistence and for the purpose of a
quantitative comparison with Ref.
\cite{pan93}, we reset the five constants in our GBD and our MDYI potentials.
For both parametrizations we require that $\sigma=12/11$, E/A = $-$16 MeV,
$\rho_{0}=0.15\ fm^{-3}$, $U(\rho,p \rightarrow \infty)\ =\ 30.5\ {\rm MeV}$,
and $m^{*}/m$ = 0.67. We then obtain K = 210 MeV, for both potentials.
We call these the new MDYI (NMDYI)
and new GBD (NGBD), respectively, to
distinguish these new parameter sets from the previous ones. We further note
that both NGBD and NMDYI give  a similar excellent  fit of the high energy
optical
potential (defined at saturation density), a desirable and important feature.

If one neglects the  momentum--dependent term, which means  C=0, the
mean field is a function of the nuclear density $\rho$ alone. This
simple Skyrme parametrization has the form (making the $\vec{r}$ dependence
implicit):
\begin{equation}
U(\rho)=A\ (\frac{\rho}{\rho_{0}})+B\ (\frac{\rho}{\rho_{0}})^{\sigma}\ .
\end{equation}
We may thus further define two additional parameter sets. The first is a
Hard potential (K~=~373~MeV) and the second a Soft potential (K~=~200 MeV).
The parameters for the GBD, MDYI, NGBD, NMDYI, and momentum--independent
Hard and Soft potentials are summarized in Table 1, together with a hard
MDYI potential(HM), which has K = 373 MeV. Note that for all those potentials
$P ( \rho_0 , T = 0 ) = 0$ and $E/A (\rho_0 , T = 0)\ =\ -16$ MeV.

Fig.~\ref{gmpoten} shows the difference between the NGBD and NMDYI
single--particle
potentials. Both those potentials produce the same bulk nuclear matter
properties at equilibrium.
We plot the potentials as a function of wave
vector $k$, for densities ranging from 0.1 to 0.5 $fm^{-3}$, in units of 0.1
$fm^{-3}$.  Both potentials have a somewhat similar momentum dependence,
but for higher densities  the NGBD is
more attractive at values of $k \alt  k_{\rm F}$ and notably more
repulsive at $k \agt k_{\rm F}$. This behaviour has been noted previously
\cite{gmw88}. We thus insist on the following important fact: even though the
two interactions have an identical high--momentum behaviour  for $\rho$ =
$\rho_0$, the similarity in their asymptotic values is not guaranteed for
densities other than equilibrium nuclear matter density.
The influence on the collective observables will be discussed in the
following sections. As an additional comparison,
we also show the momentum dependence of the GBD and NGBD
parametrizations in Fig. \ref{rgpoten}.  The two parametrizations yield almost
identical compressibilities (c.f. Table 1), but the high momentum NGBD
is much more repulsive owing mainly to its asymptotic optical potential:
$U(\rho,p\rightarrow \infty)$.  Also comparing to Wiringa's microscopic
calculations~\cite{rbw88} one realizes that NMDYI is very close in behaviour to
that of the UV14 + UVII interaction, over a wide range of momenta and
densities.  On the other hand, the high momentum part of NGBD reaches values
closer to that of the UV14 + TNI potential.  In addition to fitting
nuclear matter properties, the potentials described by Wiringa can
reproduce nucleon-nucleon
scattering and few--body data.

\section{Transverse momentum}

One important technique proposed to quantify the flow of nuclear matter is
the transverse momentum analysis~\cite{pd85}.  This method has also been
used to clarify the transverse momentum generating features of
different nuclear mean fields in the BUU approach to nucleus--nucleus dynamics.
In the framework of such studies it has been shown that under certain
circumstances, a soft
momentum--dependent potential can
produce  about the same transverse momentum as that of a hard
momentum independent--interaction~\cite{gale87,ja87}.
An effort to understand this was made in
reference~\cite{gale90}.  In order to further highlight the behaviour in a
dynamical situation of the Hard,
GBD, NGBD and NMDYI potentials, we plot in Fig.~\ref{nbbuu} the
time evolution of the average transverse momentum
for  a symmetric Nb + Nb  collision at projectile kinetic energy E =
400 MeV/nucleon at an impact
parameter b~=~2.1~fm. A sizeable
difference in the saturated transverse momentum is observed.  The hard
momentum--independent potential follows the behavior of the soft
momentum--dependent one quite closely, at this impact parameter. The asymptotic
values of their average transverse momentum are only 4 MeV/c apart. We comment
on the
behaviour of the momentum--dependent interactions below.

By turning off the hard nucleon-nucleon collisions, one can study the Vlasov
behaviour of the Hard, GBD NGBD and NMDYI potentials. From Fig. 4 one realizes
that the momentum--dependent single particle potentials alone can generate
large
transverse momenta, whereas the Hard potential can only yield very small
transverse momenta. These results are
similar to the results of Ref.~\cite{gale90}. Comparing Figs.~\ref{nbbuu} and
\ref{nbvlasov}, we can further deduce another important fact: the role of hard
two-body collisions is quite different, depending on whether the nuclear mean
field is momentum--dependent or not. Comparing the Hard and NMDYI potentials,
the fraction of the net average transverse momentum generated by adding the
collision term to the Vlasov equation is $\approx$ 100 \% and $\approx$ 42 \%,
respectively. However, it is important to point out that the transverse
momentum is generated by the nuclear mean field and the hard two--body
collisions in a highly non--linear fashion. Fig.~\ref{nbvlasov} also
tells us that, even though the GBD and NGBD potentials
have the same functional dependence on momentum and almost identical
compressibilities, they produce net transverse momenta that are very different
from each other. As discussed in section II, this result can be understood
simply in terms of the different asymptotic values of the respective
one--body potentials. Continuing our interpretation of the results in
Fig.~\ref{nbvlasov}, we find the following interesting fact: the NGBD and NMDYI
potentials produce average transverse momenta in the Vlasov mode that differ by
$\approx$ 10 MeV/c. Both these parametrizations share
the same $U (\rho_0 , \infty )$ and $K$. As mentioned previously, fitting the
static nuclear
matter properties and optical potential is not enough to predict unambiguously
the consequences of the different interactions in nonequilibrium situations. It
is also likely that realistic cases will also carry the added complication that
generally, $< \vec{p} >\ \neq\ 0$ in the GBD formulation of the one--body
potential.

Fig.~\ref{nbpy} shows the average
in--plane transverse momentum, calculated in the BUU model,  as a function
of centre of mass rapidity.  From
this figure, it is also clear that NGBD is more repulsive than NMDYI
\cite{gmw88}. We will discuss later which interaction we favor, from a
theoretical point of view.

\section{Comparison With Data}

In this section, we compare BUU calculations with experimental
data. We will first concentrate on values of the flow parameter $F$ and
transverse momentum distributions, as measured
in asymmetric heavy ion reactions by the DIOGENE
collaboration~\cite{md89} and by the
Riverside/GSI/LBL Streamer chamber group \cite{db92}.
The flow parameter $F$ is defined as
\begin{equation}
F=\ \left[ {d<P_{x}/m> \over dy} \right]_{y = y_{0}}\ .
\label{flow}
\end{equation}
Here $<P_{x}>$
is the average value of the transverse momentum
projection on the reaction plane and $y_{0}$ is the rapidity at the
intercept: $<P_{x}> \mid_{y_{0}}=0$.    Since the
 experimental efficiency cuts influence the observables, corresponding
restrictions have to be applied to the theoretical calculations in order to
compare with measured values.

Following Ref. \ref{panref}, we first turn to measurements by the
DIOGENE collaboration. There, the laboratory polar angle of the
particles is limited by
\begin{equation}
20^{\circ} \leq \theta \leq 132^{\circ}\ .
\label{angle}
\end{equation}
The transverse momentum
$P_{\perp}$ of the particles have to satisfy
\begin{equation}
P_{\perp}/m > 0.36 + 0.72 y \hspace{.45in}  {\rm if}\  y<0
\label{relapx1}
\end{equation}
\begin{equation}
P_{\perp}/m \geq 0.36 - 0.8 y \hspace{.45in} {\rm if}\  y \geq 0\ .
\label{relapx2}
\end{equation}
The measurements of rapidity distribution for ``pseudo--protons'', around
$y_{0}$ are plotted in
Fig.6 (a) for Ar + Pb at 400MeV/nucleon
at an impact parameter b=4.5 fm. Well-known geometrical arguments are used to
estimate the impact parameter \cite{imprm}.  The data shows a linear
rapidity dependence of $<P_{x}>$, in the interval [0, 1].  The flow
 parameter F is obtained by fitting the data to a straight line in the
appropriate interval, as
shown on Fig.~\ref{arpbpy45}.   The BUU calculations
were performed with 120 parallel simulations
to minimize numerical fluctuations, and with free space nucleon-nucleon
scattering cross sections.
%The effective n-n cross sections in the dense nuclear
%matter is also a very important issue to be cleared
%up.
We note in passing that since the two-body collisions contribute more to the
transverse  flow with a momentum--independent potential than with a
momentum--dependent potential as discussed previously, this observable is not
expected to be greatly sensitive to reasonable variations in the in--medium
cross
sections. To illustrate this point, compare also Fig.~\ref{nbbuu} with
Fig.~\ref{nbvlasov}.

Our simulation results  also show that the transverse momentum
$<P_{x}>$depends linearly on the rapidity around $<P_{x}>$=0
with both NGBD and NMDYI potentials. Fig.6 clearly shows
this. The fit to the experimental data is quite good with both interactions. To
increase data sensitivity to the model parameters, the extracted values
of the flow parameter, $F$, are plotted as a function of the impact parameters
b in Fig.7, together with the relevant data.  We see that the overall agreement
is quite remarkable with the NMDYI potential whereas the NGBD potential gives
a larger values than the experimental measurements. The momentum--independent
potential fails completely to reproduce the data.
Our calculation results are consistent with the results of
Ref.~\cite{pan93},  where the data is fitted by a low compressibility
GBD--type potential. We also confirm the important finding that the asymmetric
system can nicely separate out interactions of a similar compressibility but
with a different momentum dependence.  The results associated with
the hard and soft interactions of the MDYI type do not differ much in this
plot.

  Now  we turn to rapidity distributions as measured by the Streamer
Chamber.   The results of the analysis are presented in terms
of the mean in plane transverse momentum as a function of normalized
rapidity in Ar + Pb central collisions at 400 and 800
 MeV/nucleon, respectively.  All protons, whether free or bound in
clusters, have been included.  Fig. \ref{arpbpy400} shows the calculation
results of rapidity distributions with the NMDYI and NGBD potentials at
400MeV/A       in comparison with the data.   The behaviour differs
slightly from the common
S shape~\cite{doss86} due to  the asymmetry in collision geometry.
It also differs from the linear DIOGENE data as the two detectors have widely
different acceptances.
In the calculation, the maximum impact parameter was
evaluated within  a geometrical clean cut model. % and the
%integrations over impact parameter were appropriately
% weighted before being combined.
The value of $b_{max}$ used was
5.8 fm.   Our calculations with the NMDYI potential reproduce the data very
well.
A considerably larger transverse momentum transfer was
generated by the NGBD potential simulations. In Fig. \ref{arpbpy800} we compare
the results obtained with the two potentials with data obtained with the
same projectile--target combination, at 800 MeV/A.
We reach similar conclusions as in the 400 MeV/A case.

Fig.10 presents the excitation function of the average in--plane transverse
momentum in Ar + Pb collisions.  The
average transverse momentum per nucleon is evaluated from protons with
rapidities in the c.m. system greater than 0.1, 0.15,
0.2 and 0.3 for beam energies 400, 800, 1200 and 1800 MeV/A,
respectively.  The average
BUU transverse momentum with the NGBD and hard MDYI(HM)
potentials are much larger than the data. The only good fit is provided by
the NMDYI potential.    There, the agreement is striking at all energies.
Again, a hard momentum--independent potential is completely ruled out by this
data.

Summarizing this section so far, we reproduce both the DIOGENE and the Streamer
Chamber measurements quite well in terms of BUU microscopic
simulations with the MDYI--type  momentum--dependent potential, with a
compressibility K = 210 MeV. A GBD--type  momentum--dependent
potential with the same K value is not so successful and a
momentum--independent
interaction fails completely. Our findings support those of
Ref~\ref{panref}: the flow parameter data for asymmetric systems is quite
efficient in separating interactions that are
momentum--dependent from those that are momentum--independent, even though
their compressibility coefficient are the same.

We now turn to a
set of preliminary data on symmetric systems \cite{tpc}, as measured by the EOS
TPC Collaboration. Such data are of high quality, virtually free of
experimental biases. The EOS Time Projection Chamber, with its simple and
seamless acceptance, good particle identification and high statistics, was
designed to overcome the limitations of the previous generation of 4$\pi$
detectors. The Plastic Ball detector, even though having provided a seminal
contribution to the field, had a complex acceptance that was not so easily
simulated. The Streamer Chamber was somewhat limited by its particle
identification capabilities.  In the EOS TPC measurements we shall consider,
all
nuclear fragments species up to $^4$He are included. The multiplicity trigger
was set in order to select an interval centered about the value
where the flow has its maximum. This multiplicity interval corresponds to
baryon
multiplicities 0.6$M^{\rm max}\ \leq\  M\ \leq\  0.9 M^{\rm max}$. $M^{\rm
max}$ is a
value near the upper limit of the multiplicity spectrum where the height of the
distribution has fallen to half its plateau value. In our BUU calculations, we
have adjusted our impact parameter  limits to reproduce the multiplicity cuts,
in
a geometrical clean--cut model. The integration was then carried out by
sampling several impact parameter values between those two limits. The
data are in--plane transverse momentum measurements as a function of rapidity.
The flow
parameter $F$ could then be evaluated. Fig. \ref{tpcflow} shows the TPC data,
together with our calculated results. Measurements were made  for Au + Au
at beam energies 250, 400, 600, 800 and 1200 MeV/A. We display results of
calculations with a soft and a stiff momentum--dependent potential.
Calculations done with the NMDYI interaction reproduce the data exactly.

Au + Au is a reasonably large system and it might be that there exists effects
that
could safely be neglected for smaller nuclei at lower energies that are
important
here. It was brought up recently  that  an improvement in the angular momentum
conservation in the microscopic models  could perhaps lead to a re-evaluation
of the
role played by the nuclear mean field in generating transverse momentum in
heavy ion collisions \cite{ka94}. The quantitative importance of
conservation laws in microscopic models of heavy ion collisions has been
investigated before \cite{gale90b,schla91}. In the case at hand, the only
difference might
come from the fact that we are dealing here with very heavy systems at high
energies.
Thus the respective role played by two--body scattering and mean field effects
might be modified. We have considered two numerical algorithms. In the first
approach, whenever two--body scattering occurred at the microscopic level we
made sure that the direction of the reaction plane was unchanged in the centre
of mass of the colliding nucleons. The other
algorithm made sure that angular momentum was conserved {\em exactly}.
Fig.~\ref{angmom}(a) displays the effects of reaction plane conservation on the
results of BUU calculations, for Au + Au collisions at 1 GeV/A. The impact
parameter range and kinematical cuts were adjusted to match those of the TPC.
Fig.~\ref{angmom}(b) shows the consequences of reaction plane and exact angular
momentum conservation on cascade simulations of the same nuclear reaction. The
exact algorithms used are described in Ref.~\ref{gale90b}. The imposition of
exact angular momentum conservation increases the flow parameter by roughly
23\% in cascade simulations. We can also see that
the dominant effect in angular momentum
conservation comes from keeping the direction of the reaction plane constant
in individual two--body collisions. In the BUU calculations, reaction
plane constraints
raise the flow by only 8\%. The net effect on the transverse momentum can
readily by
appreciated in those two figures.  Basically, the effect on BUU calculations in
considerably smaller than in cascade approaches.

\section{Summary}
We have used the Boltzmann-Uehling-Uhlenbeck equation to describe the
dynamics of nucleus-nucleus collisions. Concentrating on the
momentum--dependent features of the one--body self--consistent nuclear mean
field, we have seen that the precise functional dependence on momentum
of the interaction was important. Taking two potentials with the exact same
characteristics at saturation density and zero temperature (NGBD and NMDYI),
we have shown that their
behaviour in situations removed from equilibrium could be quite different. From
a purely theoretical point of view we believe that approaches based on
MDYI--type interactions are on a firmer basis. In GBD--like approaches, the
quantity $< \vec{p} >$ was put in by hand to
enforce the Galilean invariance of the potential. MDYI has Galilean invariance
from the start and furthermore, the fact that it can be identified with the
Fourier transform of a Yukawa potential is pleasing. Both interactions have
the virtue of being relatively simple to handle (MDYI is however trickier to
implement numerically). Again, in equilibrium or close to equilibrium
situations it should make little difference which is used. As far as the
results outlined in this paragraph go, our investigations follow in the steps
of some of our previous works \cite{gale90}.

Furthermore, keeping in mind the quantitatively different results
obtained with NGBD and
NMDYI, we have confirmed the idea put forward by Pan and Danielewicz
\cite{pan93}.  By performing calculations to address data on symmetric and
asymmetric systems at high energies, one can indeed assess the importance of
the density--dependent and momentum--dependent terms in the nuclear equation of
state, separately. In pursuing this point, we have for the first time compared
DIOGENE and EOS TPC data with BUU results. We find that all the data we have
considered in this paper can be reproduced with a momentum--dependent
interaction
with a nuclear compressiblity coefficient of K = 210 MeV. Also, we have
verified again the importance of angular momentum conservation on the
generation of transverse momentum in high energy heavy ion collisions. Relaxing
the conservation law leads to a slight variation in the
flow parameter in BUU collisions. This change should be considered in high
precision fits of the experimental data as it should undoubtedly lead to lower
values of $\chi^2$ \cite{schlag}. This does not however alter the general
conclusions reached in this work.

Finally, it is worth pointing out that after more than a decade of careful
experimental investigations and theoretical progress, a consistent picture of
the behaviour of nuclear matter at high temperatures and densitites is
emerging. Perhaps the crudest way of characterizing the nuclear equation of
state is by its compressibilty coefficient and this value is now stabilizing to
some number around 210 MeV. The fact that low and high energy heavy ion
experiments seem to require compatible values of K is satisfying. The fact
that high quality, bias--free, exclusive experimental  is now available and
will continue to be generated in the immediate future will set even more
stringent tests for the models.  Some challenging problems remain to be
successfully tackled in theory. For example, the area of composite
production is an active area of research \cite{danbe91}. In the context
of nuclear flow, this is
a pressing issue as it now clear that composites bear the greatest sensitivity
to collective behaviour in high energy heavy ion collisions.

\acknowledgments

We wish to acknowledge fruitful discussions with P. Danielewicz, D. Keane and
Q.
Pan. Our research is supported in part by the Natural
Sciences and Engineering Research Council of Canada and in part by the FCAR
fund
of the Qu\'ebec government.

\clearpage
%----------- references ----
%\begin{thebibliography}{24}

%\end{thebibliography}
\clearpage
\widetext
\begin{table}
\caption{We write here the parameters and characteristics of the
single--particle potentials we have introduced in the main text.
\label{tapa}}
\vspace{1cm}
\begin{tabular}{ccccccccccc}
Model &A &B & $\sigma$ &C &$\Lambda$ &$m*/m$ &$U(\rho_{0},p_{\rm F})$
&$U(\rho_{0},0)$ &$U(\rho_{0}, \infty)$ &K    \\
&(MeV) &(MeV) & &(MeV) &(MeV) & &(MeV) &(MeV) &(MeV)
 & (MeV)  \\
\tableline
Soft  &$-$351.3 & 300   & 7/6    &  0     &         &   1    & $-$51.3
   &  $-$51.3        & $-$51.3
                    &  200   \\
Hard  &$-$120.5 & 69.2  & 2      &  0     &         &   1    & $-$51.3
   &  $-$51.3        & $-$51.3
                    &  373    \\
GBD   &$-$144.9 & 203.3 & 7/6    & $-$75    &  400    &   0.7  & $-$53.3
   &  $-$76.3        & $-$1.34
                    &  215    \\
MDYI  &$-$110.44& 140.9 & 1.24   & $-$64.95 &  415.7  &   0.67 & $-$52.9
   &  $-$75          & 30.5
                    &  215    \\
NGBD  &$-$227.5 & 347.7 & 12/11  & $-$103.9 &  495.4  &   0.67 & $-$51.4
   &  $-$73.5        & 30.5
                    &  210     \\
NMDYI &$-$322   & 352.5 & 12/11  & $-$62.75 &  417    &   0.67 & $-$51.4
   &  $-$72.4        & 30.5
                    &  210  \\
HM& $-$9.0 & 39.5 & 2.27 & $-$62.75 & 417 & 0.67 & $-$51.4 & $-$72.4 &
30.5 & 373
\end{tabular}
%\end{array}
\label{table1}
\end{table}
\clearpage
\mediumtext
%figure 1
\begin{figure}
\caption{A comparison of the momentum--dependent NMDYI and
NGBD potentials, adjusted to produce identical
bulk properties in cold nuclear matter.  The abscissa shows the wave
number. Starting from the bottom,
the different curves are for densities of 0.1, 0.2, 0.3, 0.4 and 0.5
$fm^{-3}$.}
\label{gmpoten}
\end{figure}

%figure 2
\begin{figure}
\caption{Same caption as in Fig. 1, but with the GBD and NGBD potentials.}
\label{rgpoten}
\end{figure}
%figure 3

\begin{figure}
\caption{Average in plane transverse momentum per nucleon versus time
for BUU calculations
of  Nb + Nb collisions at 400 MeV/nucleon, at an impact parameter b = 2.1 fm.
The results are for the Hard, GBD, NGBD
and NMDYI potentials.}
\label{nbbuu}
\end{figure}

%figure 4
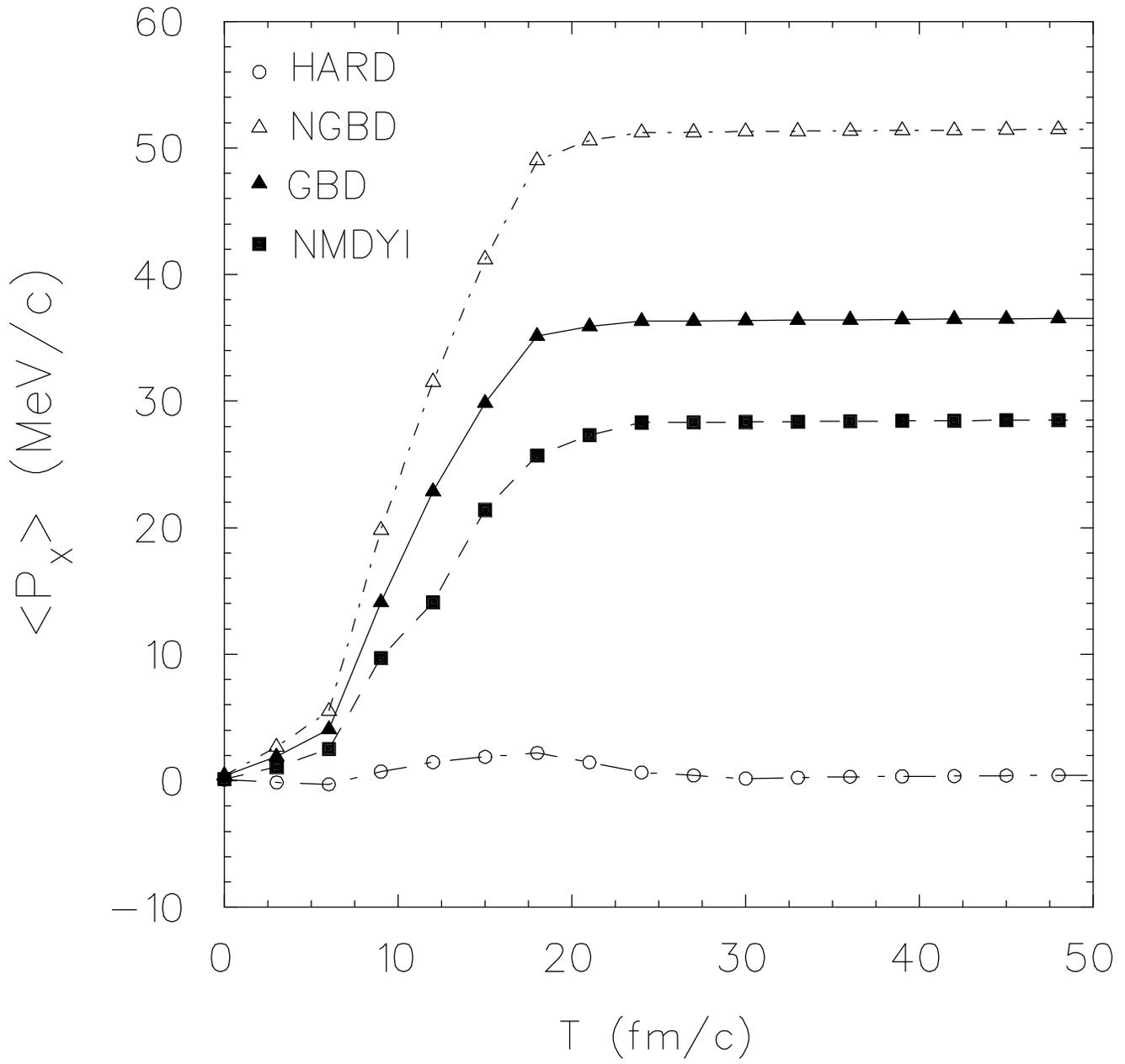
\begin{figure}
\caption{Average in plane transverse momentum per nucleon versus time for
Vlasov
calculations of Nb + Nb at 400 MeV/nucleon, at an impact parameter b = 2.1 fm.
The results are for the Hard, GBD, NGBD and NMDYI potentials.}
\label{nbvlasov}
\end{figure}

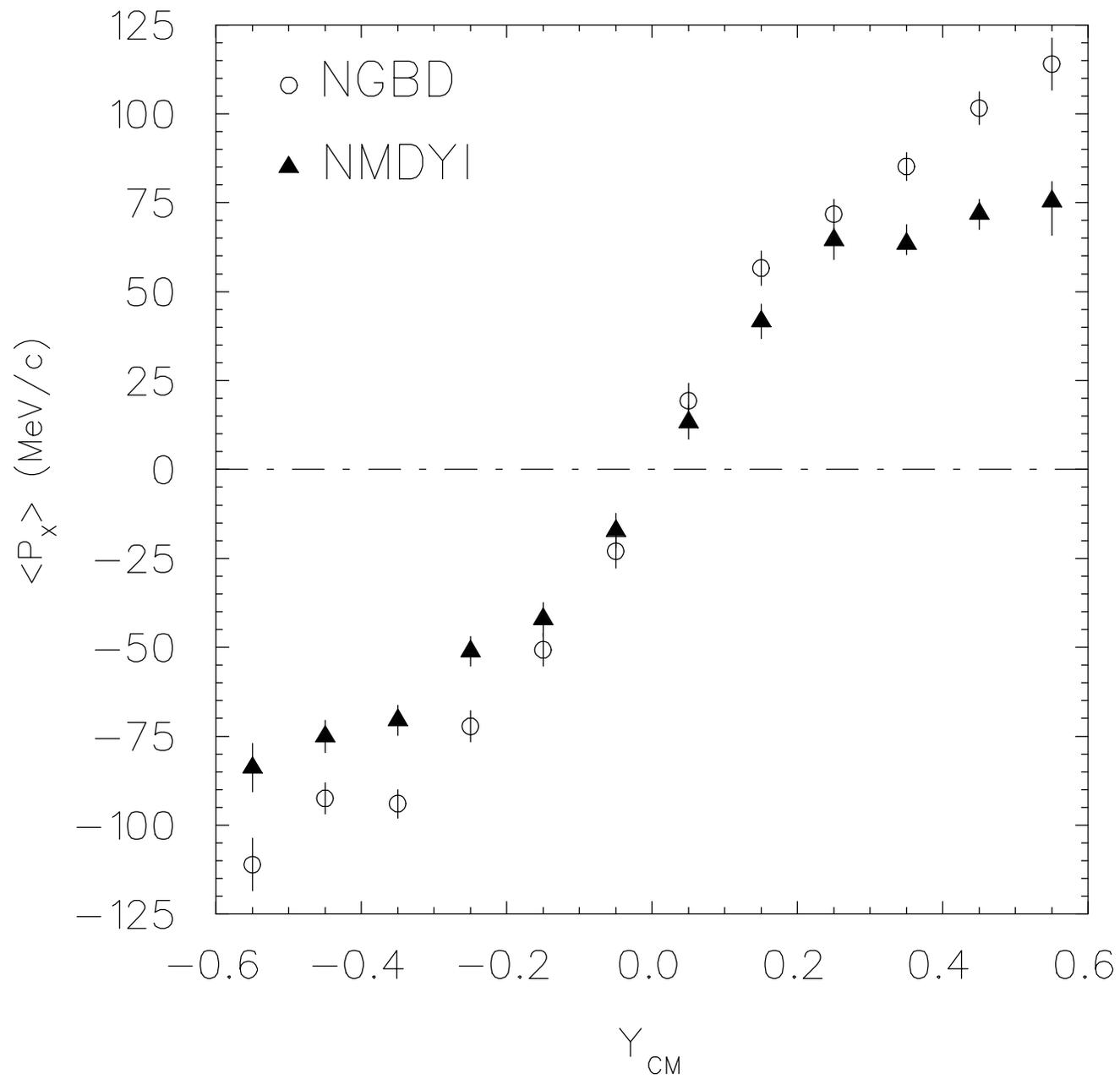
\begin{figure}
%figure 5
\caption{Average in plane transverse momentum distributions versus
centre of mass rapidity for Nb + Nb
for b = 2.1 fm and beam energy 400 MeV/nucleon. The
results are for the NGBD and NMDYI interactions.}
\label{nbpy}
\end{figure}

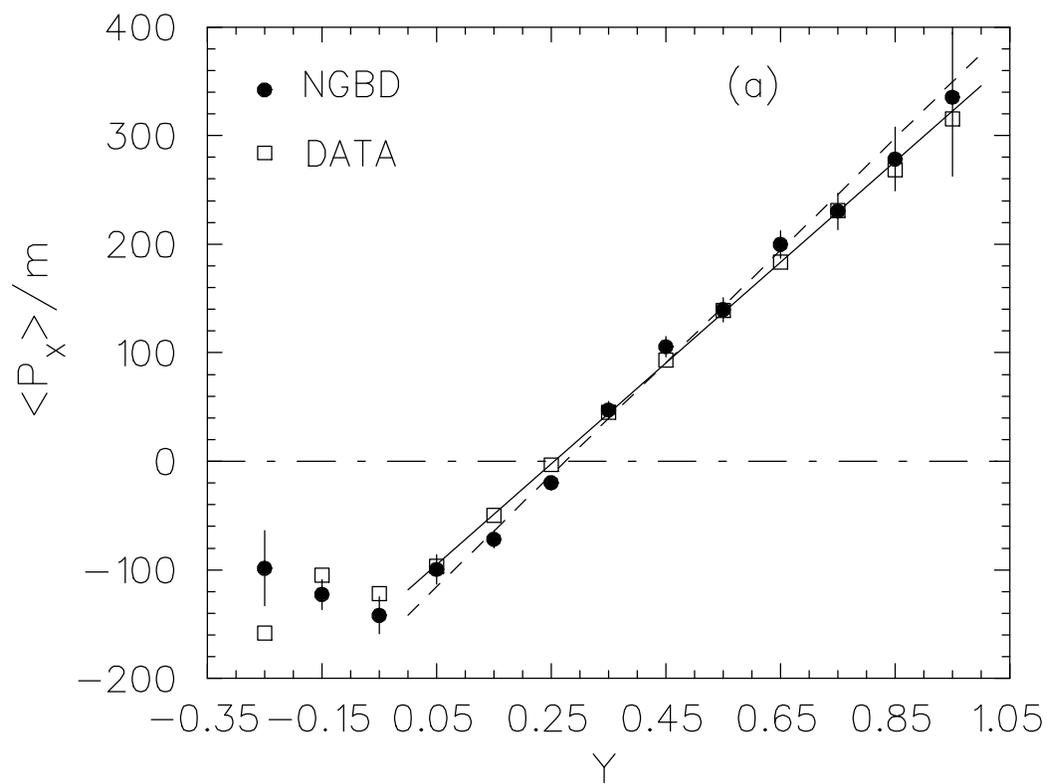
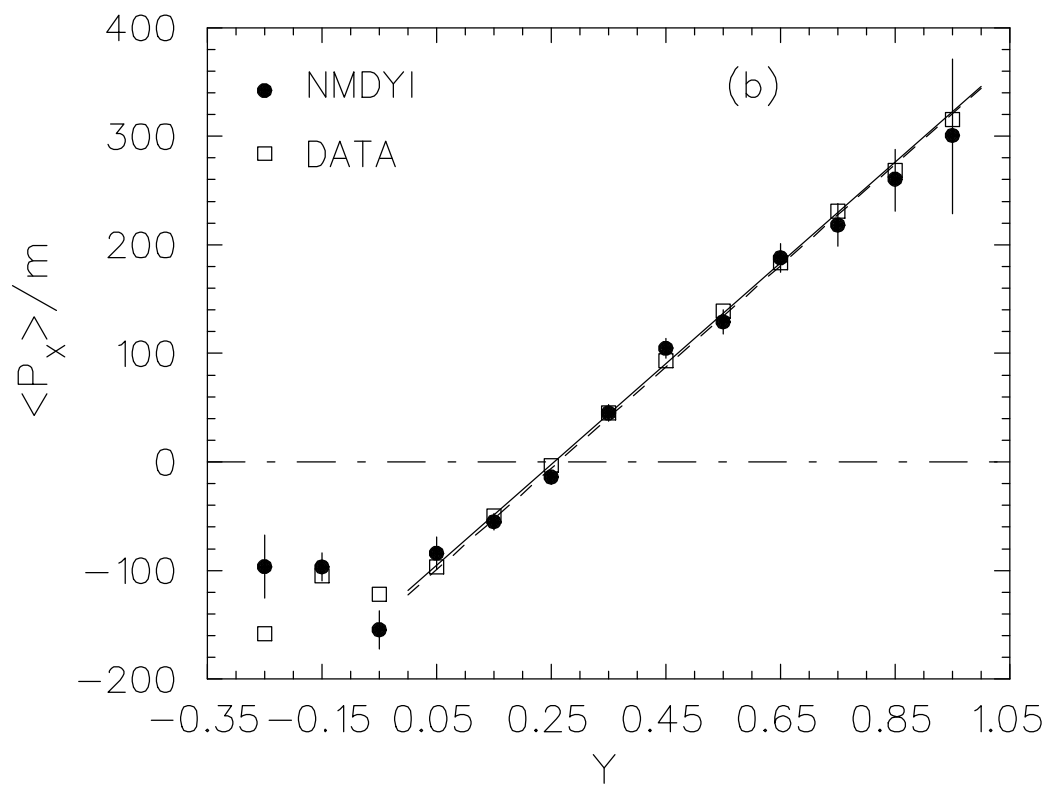
\begin{figure}
%figure 6
\caption{Average in plane transverse momentum (divided by the proton
mass) as a function of rapidity in the Ar + Pb
reactions at 400 MeV per projectile nucleon at an impact parameter b=4.5
fm.  The solid and dashed lines represent linear fits through data [19] and
calculation, respectively.}
\label{arpbpy45}
\end{figure}

%figure 7
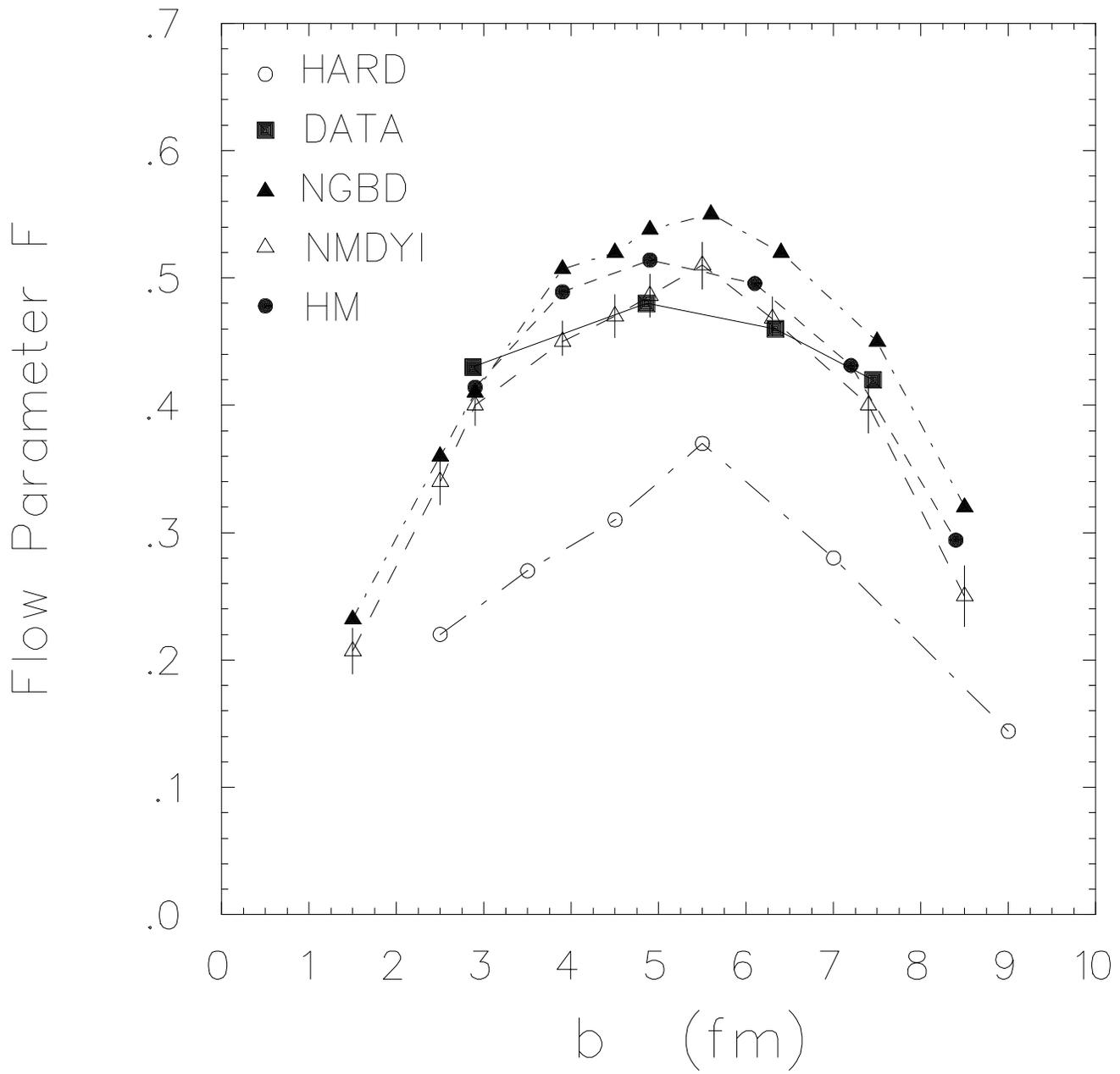
\begin{figure}
\caption{Impact parameter dependence of the flow parameter $F$ for Ar+Pb
reactions.  The results of BUU calculations with different single particle
potentials, Hard, NGBD, NMDYI and HM,  are compared with the data of Ref. [19].
Error bars
in the theory reflect statistical errors and are only given for one set
of calculations.}
\label{arpbfb}
\end{figure}

%figure 8
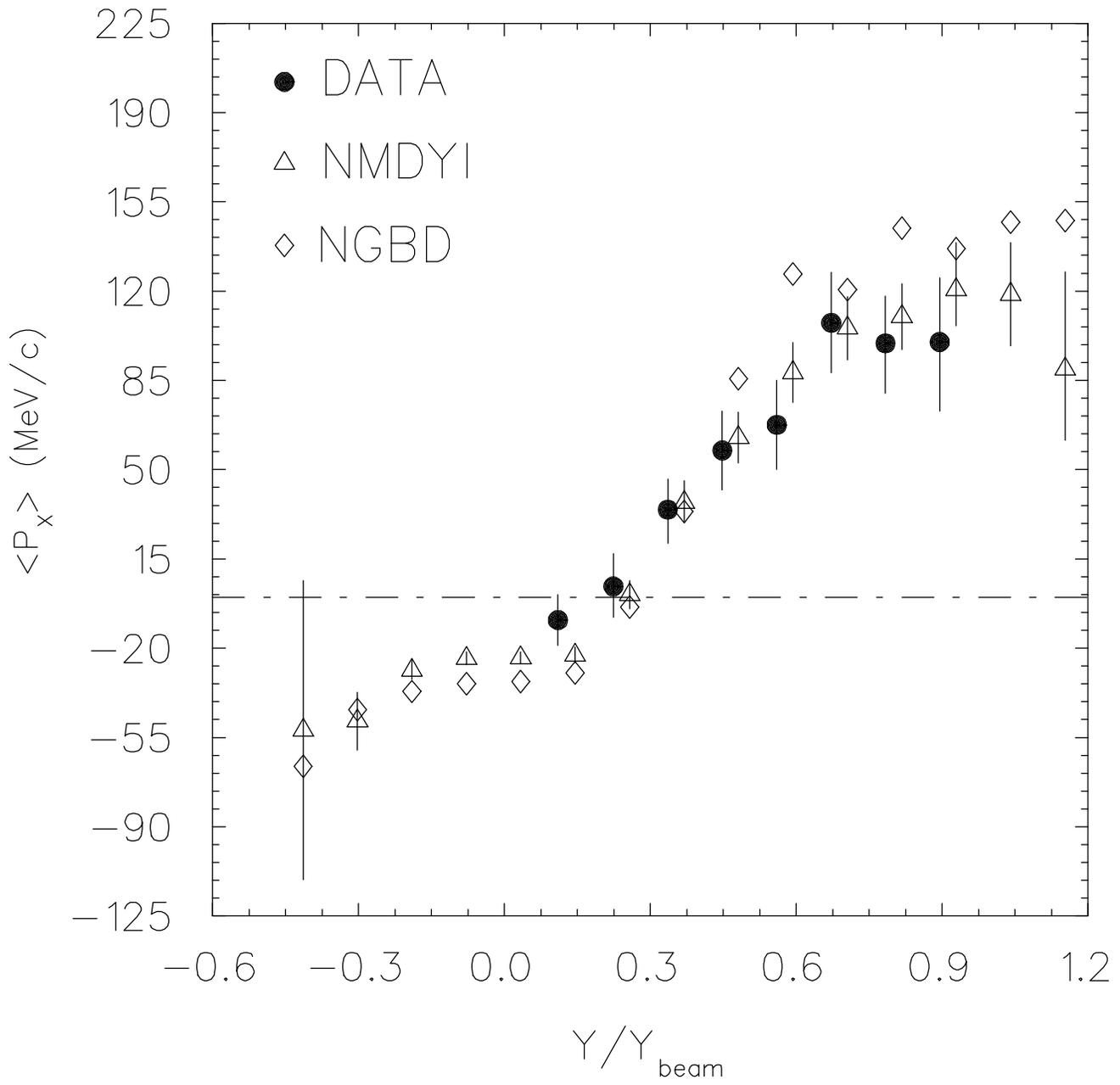
\begin{figure}
\caption{Average in plane transverse momentum
as a function of normalized rapidity
in central Ar + Pb collisions at 400 MeV per projectile nucleon.
The data of Ref. [20] are compared with
BUU calculations with the NMDYI and NGBD  potentials. Errors bars in the theory
reflect statistical errors only and are given for one set of calculations.}
\label{arpbpy400}
\end{figure}

%figure 9
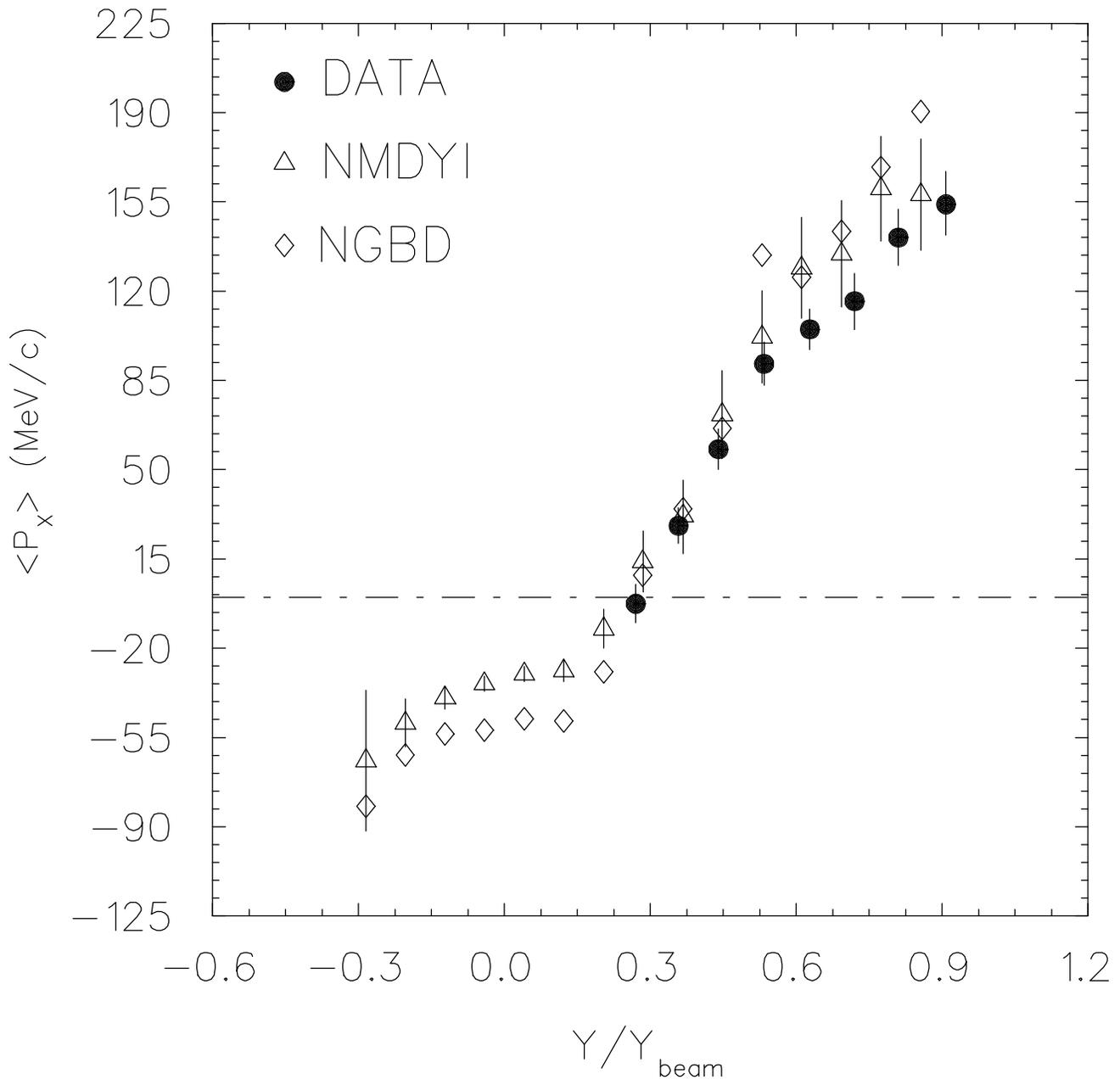
\begin{figure}
\caption{Same caption as in Fig. 8 but with incident kinetic energy 800
MeV/A.}
\label{arpbpy800}
\end{figure}

%figure 10
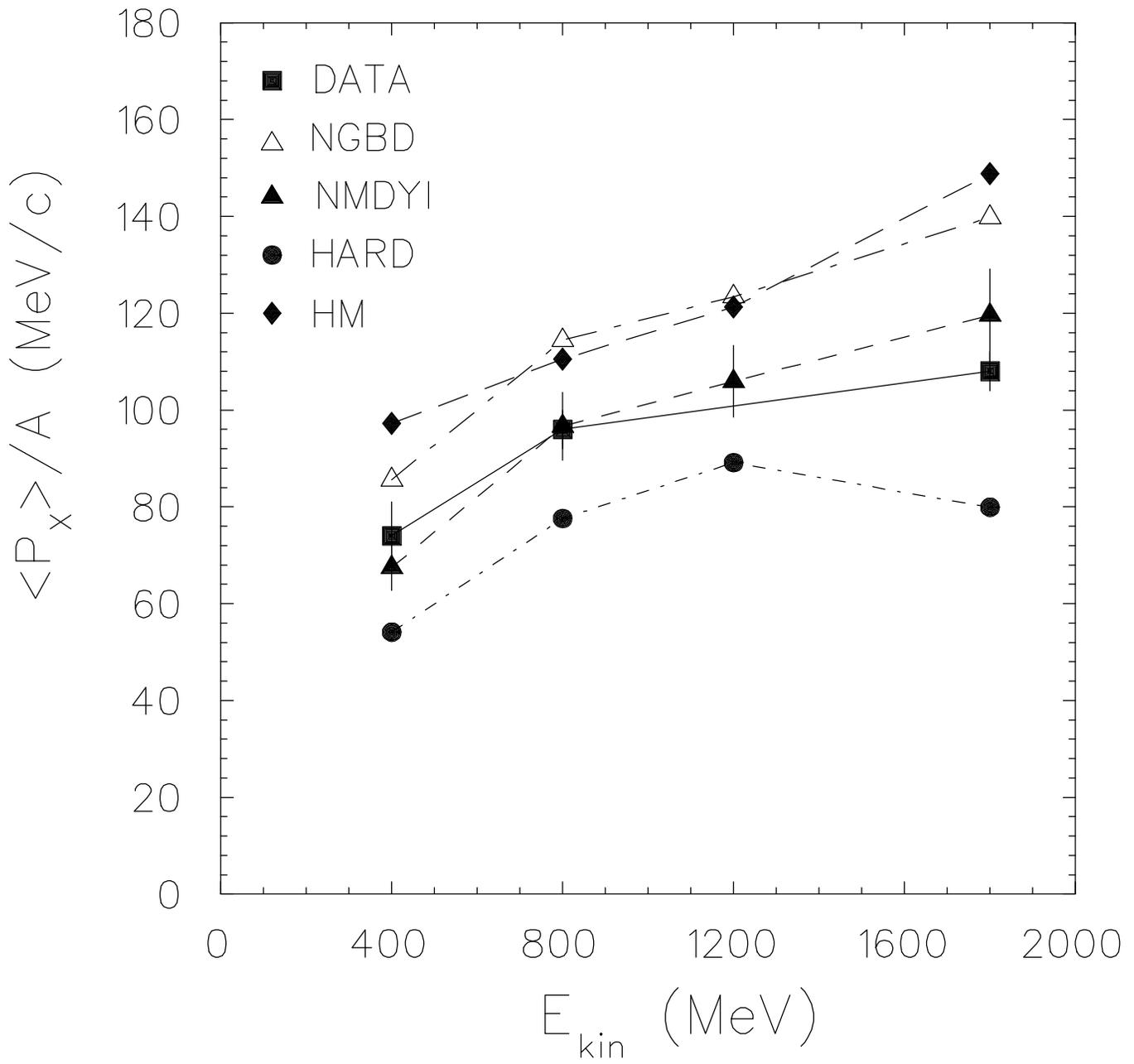
\begin{figure}
\caption{We plot the excitation function of the average transverse momentum
per nucleon in the reaction plane for the forward centre of mass hemisphere
as a function of beam energy for Ar+Pb reactions.  The data of
Ref. [20] are compared with the BUU
calculations with the Hard, HM, NGBD and
NMDYI potentials. Errors bars in the theory reflect statistical errors only and
are given for one set of calculations.}
\label{arpbpe}
\end{figure}

%fig. 11
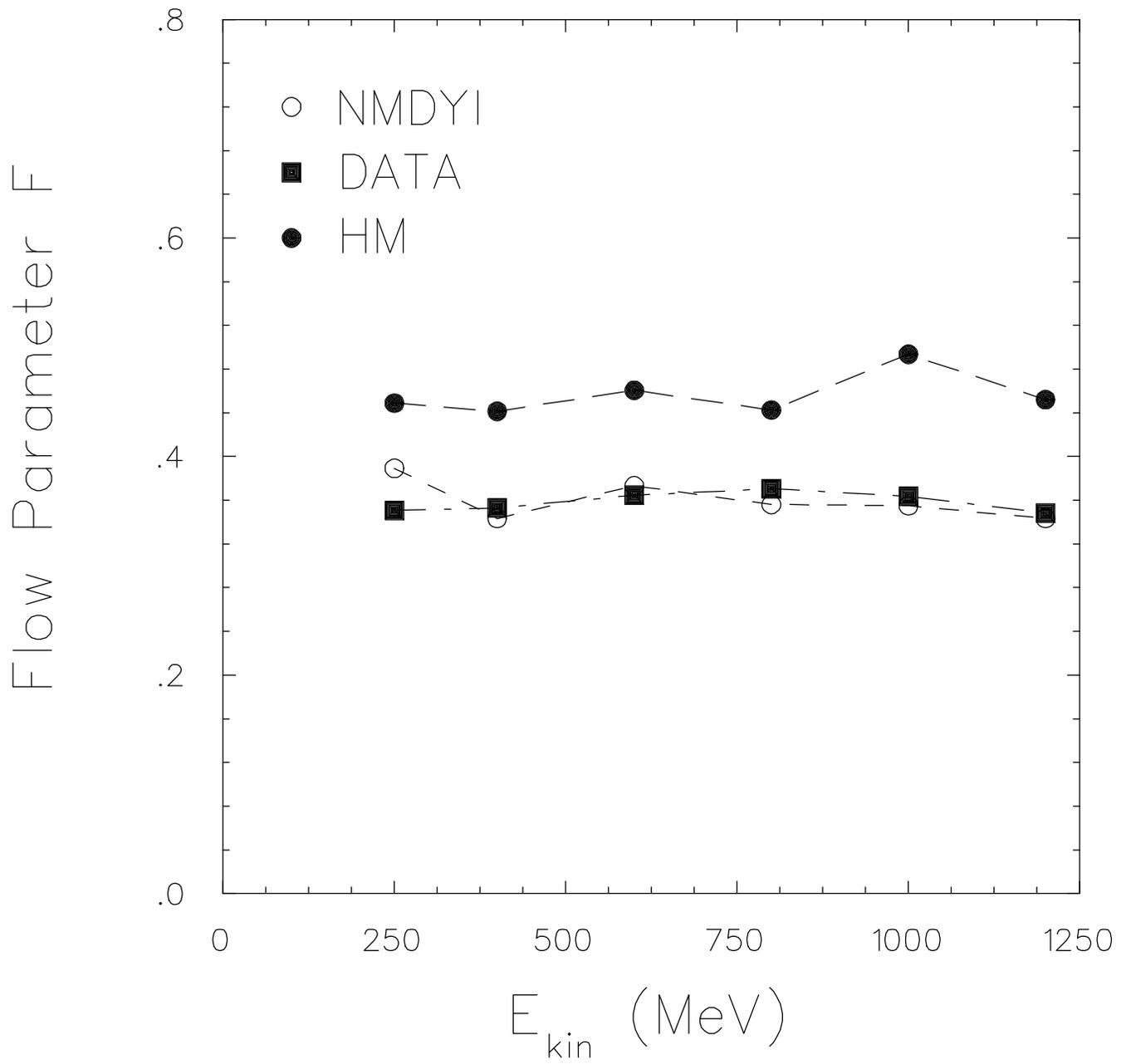
\begin{figure}
\caption{We plot the excitation function of the flow parameter $F^\prime\ =\
F \times  y_{\rm beam}^{\rm cm}$, where $F$ is defined
in the main text. The solid squares refer to Au + Au data as measured by the
EOS TPC
Collaboration [21], the circles are calculations done within the BUU approach,
with
a soft and stiff compressibility coefficient. The numerical uncertainties in
the calculations are of the order of 10\%, as previously.}
\label{tpcflow}
\end{figure}

%fig. 12
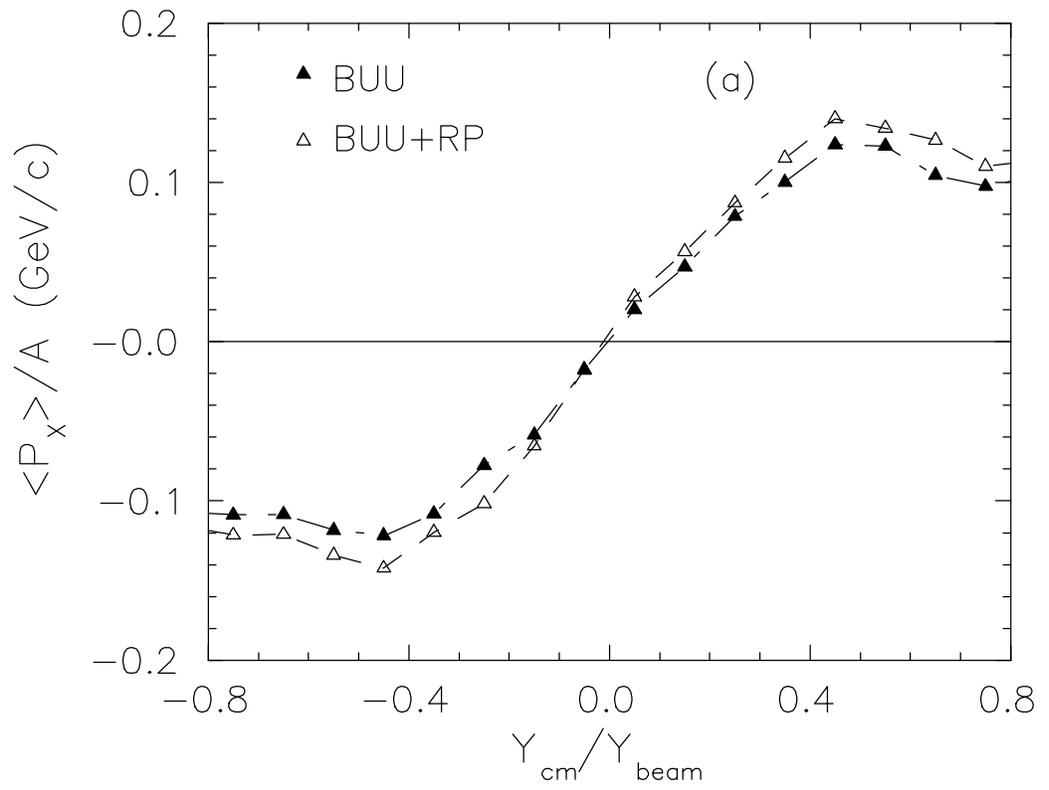
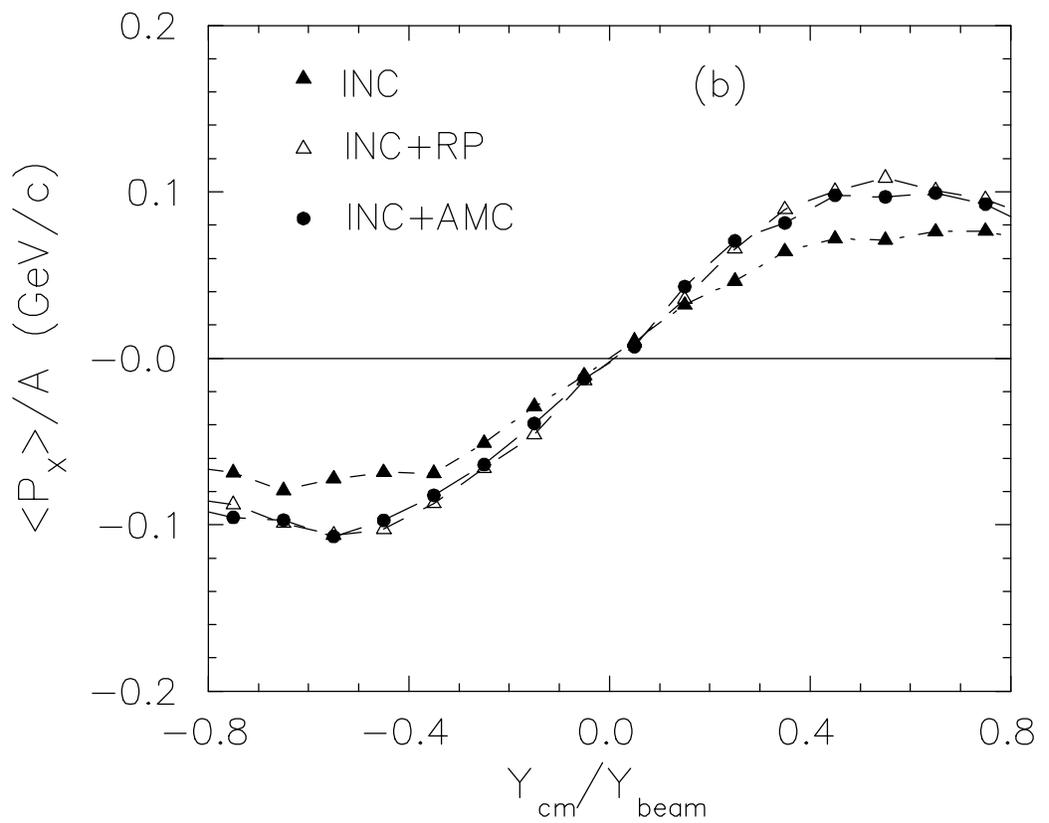
\begin{figure}
\caption{We plot the transverse momentum generated in Au + Au collisions at
1 Gev/A against rapidity in the centre of mass. We investigate the consequences
of (a) imposing a reaction plane (RP) on each two--body collision in the BUU
model
and (b) reaction plane (RP) and exact angular momentum conservation (AMC)
in a cascade approach.}
\label{angmom}
\end{figure}

\end{document}